\documentclass[12pt,a4paper]{article}    

\usepackage{graphicx,amsfonts}

\newcommand{\be}{\begin{equation}}
\newcommand{\ee}{\end{equation}}        

\newcommand{\bea}{\begin{eqnarray}}
\newcommand{\eea}{\end{eqnarray}}

\newcommand{\ba}{\begin{align}}
\newcommand{\ea}{\end{align}}  

\newcommand{\s}{{}^* \negthinspace}

\newcommand{\M}{{\mathcal{M}}}
\newcommand{\U}{{\mathcal{U}}}

\newcommand{\Nbd}{\mathcal{N}}

\newcommand{\R}{\ensuremath{\mathbb{R}}}

\newcommand{\lra}{\Leftrightarrow}

\newcommand{\smallfrac}[2]{\mbox{\small $\frac{#1}{#2}$}}

\renewcommand{\vec}{\textbf}
\newcommand{\tns}{\textbf}

\newcommand{\text}[1]{\mbox{#1}}

\newcommand{\qed}{\rule{3mm}{3mm}}

\newcommand{\implies}{\Rightarrow}

\newcommand{\thisdoc}{paper }
\newcommand{\thisdocthesis}{paper }

\newenvironment{proof}[0]{\noindent\textbf{Proof:}\\}{\hfill\qed}

\newtheorem{thm}{Theorem}[section]
\newtheorem{defn}{Definition}[section]

\newtheorem{prop}{Proposition}[section]

\newtheorem{cor}{Corollary}[section]

\begin{document}

\title{General properties of cosmological models
       with an Isotropic Singularity}

\author{Geoffery Ericksson\thanks{Advanced Computational Modelling Centre,
The University of Queensland,
Brisbane, QLD  4072.
E-mail: {\em {\tt gbe@acmc.uq.edu.au}}
}
\and
Susan M. Scott\thanks{Department of Physics,
Faculty of Science,
The Australian National University,
 Canberra, ACT 0200, Australia.
 E-mail: {\em {\tt Susan.Scott@anu.edu.au}}
}
}

\date{}            
\maketitle          


\abstract{ Much of the published work regarding the Isotropic
Singularity is performed under the assumption that the matter
source for the cosmological model is a barotropic perfect fluid,
or even a perfect fluid with a $\gamma$-law equation of state.
There are, however, some general properties of cosmological models
which admit an Isotropic Singularity, irrespective of the matter
source. In particular, we show that the Isotropic Singularity is a
point-like singularity and that vacuum space-times cannot admit an
Isotropic Singularity. The relationships between the Isotropic
Singularity, and the energy conditions, and the Hubble parameter
is explored. A review of work by the authors, regarding the
Isotropic Singularity, is presented. }


\section{Introduction}
The concept of an Isotropic Singularity (IS) was introduced to the
field of mathematical cosmology by Goode and Wainwright in 1985
\cite{GW85}. Since that time much effort has been directed at
investigating the physical consequences of the definition.  This
paper is devoted to an examination of some of the general
properties of cosmological models which admit an IS. Previously
published work by the authors, in this area, is reviewed in
Section \ref{sec:review}. Prior to that, for reference, the
definition of an IS is given in Section \ref{sec:defn_of_an_IS}
together with the associated definition of a cosmic time function.
A proof follows which details a freedom in the choice of the
conformal factor associated with an IS.  For cosmologies
comprising a fluid flow, the definitions of a fluid congruence
which is regular at an IS and a fluid congruence which is
orthogonal to an IS are given in Section
\ref{sec:defn_of_fluid_flow}.  When a cosmological model admits an
IS at which the fluid flow is regular, it is shown that a
conformal factor can always be chosen such that the expansion
scalar associated with the unphysical fluid flow becomes zero at a
given point in $\s\M$ (with $T\geq0$).

In Section \ref{sec:vacuum_cosmologies} the issue of whether or
not vacuum cosmologies can admit an IS is investigated.  The IS is
proven to be a point-like singularity, as one would expect, in
Section \ref{sec:singularity_type}.  In Section
\ref{sec:hubble_parameter} the Hubble parameter is shown to become
infinite at the initial singularity of any cosmological model with
an IS. The relationship between the energy conditions and the IS
for perfect fluid cosmological models is examined in detail in
Section \ref{sec:energy_conditions}.


\section{Definition of an Isotropic Singularity}
\label{sec:defn_of_an_IS}

Following the introduction of the IS concept by Goode and
Wainwright \cite{GW85},  Scott \cite{Scott88,Scott89,SSGVR}
amended their definition to remove some inherent redundancies. It
is this amended definition which is given in Definition
\ref{defn:IS} and which will be used throughout this
\thisdocthesis\hspace{-1.5mm}.

\begin{defn}[Isotropic singularity]
\label{defn:IS} A space-time $(\M ,\tns{g})$ is said to admit an
\emph{isotropic singularity}
 if there exists a space-time $(\s\M , \s \tns{g})$, a smooth cosmic time
function $T$ defined on $\s\M$, and a conformal factor $\Omega
(T)$ which satisfy
\begin{enumerate}
\item $\M$ is the open submanifold $T>0$,
\item $\tns{g}= \Omega^{2}(T) \s \tns{g}$ on $\M$, with $\s \tns{g}$
regular
(at least $C^{3}$ and non-degenerate) on an open neighbourhood of
$T=0$,
\item $\Omega(0) = 0$ and $\exists\thinspace b>0$ such that
        $\Omega \in C^{0}[0,b] \cap  C^{3}(0,b]$ and $\Omega(0,b] >0$,
\item $\lambda \equiv \lim_{T \rightarrow 0^{+}} L(T)$ exists,
        $\lambda\neq 1$, where
        $L \equiv \frac{\Omega''}{\Omega}
        {\left( \frac{\Omega}{\Omega'} \right)}^{2}$
        and a prime denotes differentiation with respect to T.
\end{enumerate}
\end{defn}

\begin{figure}[!ht]
\begin{center}
\includegraphics[]{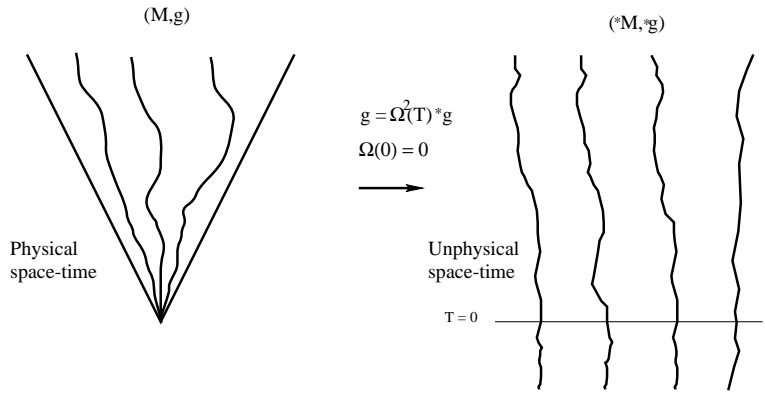}
\end{center}
\caption{Pictorial interpretation of the isotropic singularity.}
\label{is_pic}
\end{figure}

The definition of an isotropic singularity is described
pictorially in Figure \ref{is_pic}. The definition of a cosmic
time function \cite{SSGVR}\cite[p198]{HE73} which is used in
conjunction with Definition \ref{defn:IS}, is the one given in
Definition \ref{defn:cosmic_time_function}.


\begin{defn}[Cosmic time function]
\label{defn:cosmic_time_function} A function $T$ on $\s \M$ is
said to be a \emph{cosmic time function} if $\s \nabla T$ is
everywhere timelike on $\s\M$ with respect to $\s\tns{g}$.
\end{defn}

It should be noted that a cosmic time function defines a family of
spacelike hypersurfaces \{$T = $ constant\} in $\s\M$ and hence in
$\M$. A cosmic time function also increases along every
future-directed non-spacelike curve.

Since $(\M,\tns{g})$ is the space-time in hand---in our case,
typically, a cosmological solution of the Einstein field equations
(EFE)---it is usual to call it the \emph{physical space-time}. The
conformally related space-time, $(\s\M, \s \tns{g})$, is then
called the \emph{unphysical space-time}.

Although the definition of an IS, and the arrow in Figure
\ref{is_pic}, lead one to think of the unphysical space-time being
``created'' from the physical space-time, it is actually useful to
consider the situation in the reverse fashion. In this way, the
singularity in the physical space-time should be regarded as
arising due to the vanishing of the conformal factor, $\Omega(T)$,
at the regular hypersurface $T=0$ in $\s\M$.

Two noteworthy features of the definition of an IS are that it is
coordinate-independent, as well as being independent of the EFE,
and hence of the source of the gravitational field.

The conformal factor in the definition of an IS is not unique, as
will be proven in the following Proposition.

\pagebreak

\begin{prop}
\label{prop:conformal_factor_freedom} The conformal factor
associated with a particular IS can be multiplied by a factor of
$e^{k(T)}$,  \be \text{i.e.,\ } \tilde{\Omega}(T) =
e^{k(T)}\Omega(T),
\end{equation}
where $k(T)$ is a $C^3$ function of $T$ on $\R$, and still satisfy
the conditions of the definition of an IS.
\end{prop}

\begin{proof}
An outline of this proof was given by Goode and Wainwright
\cite{GW85}. If the space-time $(\M,\tns{g})$ admits an IS, then
there exists a conformally related space-time $(\s\M,\s\tns{g})$
which satisfies the conditions of Definition \ref{defn:IS}.  We
now create a second unphysical space-time
$(\s\M,\s\tilde{\tns{g}})$ from the given unphysical space-time
$(\s\M,\s\tns{g})$ by using the same manifold $\s\M$ and a metric
$\s\tilde{\tns{g}}$ which is conformally related to $\s\tns{g}$ by
the conformal factor $e^{k(T)}$, where $k(T)$ is a $C^3$ function
of $T$ on $\R$, i.e., $ \s\tns{g} = e^{2k(T)}\s\tilde{\tns{g}}$.
Note that the cosmic time function $T$ on $\s\M$ is the same for
both $(\s\M,\s\tns{g})$ and $(\s\M,\s\tilde{\tns{g}})$. The metric
$\tns{g}$ of the physical space-time $(\M,\tns{g})$ is related on
$\M$ to the metric $\s\tilde{\tns{g}}$ of the second unphysical
space-time $(\s\M,\s\tilde{\tns{g}})$ by the conformal factor
$\tilde{\Omega}=e^{k(T)}\Omega$, i.e., $\tns{g}=
\tilde{\Omega}^2(T)\s\tilde{\tns{g}}$. Figure \ref{fig:theta_zero}
shows the relationship between the physical space-time and the two
unphysical space-times.

\vspace{5mm}
\newcommand{\ymax}{65}
\newcommand{\xmax}{170}
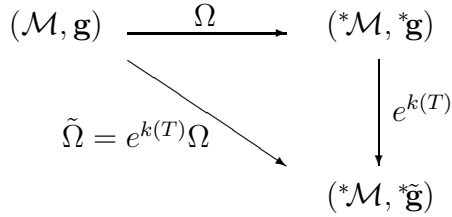
\begin{figure}[h]
\begin{center}
\begin{picture}(\xmax,\ymax)
\put(0,\ymax){$(\M,\tns{g})$} \put(120,\ymax){$(\s\M,\s \tns{g})$}
\put(120,0){$(\s\M,\s\tilde{\tns{g}})$}
\put(45,\ymax){\vector(1,0){60}} \put(45,55){\vector(3,-2){60}}
\put(140,55){\vector(0,-1){40}}
\put(70,68){$\Omega$} \put(145,32){$e^{k(T)}$}
\put(20,22){$\tilde{\Omega}=e^{k(T)}\Omega$} 
\end{picture}
\caption{Relationship between the physical space-time and the two
unphysical space-times.} \label{fig:theta_zero}
\end{center}
\end{figure}

Entities that exist in the second of the unphysical space-times,
$(\s\M,\s\tilde{\tns{g}})$, will be denoted by a star and an
over-tilde, e.g., $\s \tilde{A}$.

We now check each of the conditions outlined in Definition
\ref{defn:IS}.
\begin{enumerate}
\item $\M$ is still the open submanifold $T>0$ of $\s\M$.
\item $\tns{g} = \tilde{\Omega}^2(T)\s\tilde{\tns{g}}$ on $\M$, where
$\s\tilde{\tns{g}}= e^{-2k(T)}\s\tns{g}$.  Since $\s\tns{g}$ is at
least $C^3$ and non-degenerate on an open neighbourhood of $T=0$,
and $e^{-2k(T)}$ is a $C^3$ positive function of $T$ on $\R$, it
follows that $\s\tilde{\tns{g}}$ is at least $C^3$ and
non-degenerate on an open neighbourhood of $T=0$.
\item The relationship between the conformal factor $\Omega$ and
the conformal factor $\tilde{\Omega}$ is given by $\tilde{\Omega}
= e^{k(T)}\Omega$.  We know that $\Omega(0) = 0$ and
$\exists\thinspace b>0$ such that $\Omega \in C^{0}[0,b] \cap
C^{3}(0,b]$ and $\Omega(0,b] >0$.  Clearly $\tilde{\Omega}(0) =
0$, and since $e^{k(T)}$ is a $C^3$ positive function of $T$ on
$\R$, $\tilde{\Omega} \in C^{0}[0,b] \cap C^{3}(0,b]$ and
$\tilde{\Omega}(0,b] >0$.
\item \bea
\tilde{L}(T) &=&
    \frac{\tilde{\Omega}''\tilde{\Omega}}{(\tilde{\Omega}')^2} \\
     &=& \frac{\Omega''\Omega+2\Omega'\Omega k' + \Omega^2 k''
         + \Omega^2 (k')^2}
              {(\Omega'+\Omega k')^2}\\
     &=& \frac{ \frac{\Omega''\Omega}{(\Omega')^2} +
        2k'\frac{\Omega}{\Omega'} +
        k''\left(\frac{\Omega}{\Omega'}\right)^2 +
        (k')^2\left(\frac{\Omega}{\Omega'}\right)^2
        }{[1+\frac{\Omega}{\Omega'}k']^2}
\end{eqnarray}
It was shown by Scott \cite{SSGVR} that $\lim_{T\rightarrow
0^+}\frac{\Omega}{\Omega'} = 0$.  Now $k(T)$ is a $C^3$ function
of $T$ on $\R$, and so $k'\frac{\Omega}{\Omega'}\rightarrow 0$ and
$k''\left(\frac{\Omega}{\Omega'}\right)^2\rightarrow0$, as
$T\rightarrow 0^+$. It follows that \be \tilde{L}(T)=
\frac{\Omega''\Omega}{(\Omega')^2}[1+o(1)] \quad\text{as
}T\rightarrow 0^+,
\end{equation}
and therefore \be \lim_{T\rightarrow 0^+}\tilde{L}(T)=
\lim_{T\rightarrow 0^+}L(T)=\lambda \quad(\neq 1).
\end{equation}
\end{enumerate}
The conditions for $(\s\M,\s\tilde{\tns{g}})$ to be an unphysical
space-time conformally related to $(\M,\tns{g})$ have now been
satisfied.
\end{proof}


\section{Fluid flow}
\label{sec:defn_of_fluid_flow}
The definition of an Isotropic Singularity, as it stands,  allows
cosmological models to admit an ``isotropic'' singularity whose
singularities are, in no sense, actually isotropic, or
quasi-isotropic, or Friedmann-like \cite{Goode82,GW85}.  For
example, the exact viscous fluid FRW cosmology of Coley and Tupper
\cite{CT83} can be shown to have an IS \cite{GW85},  and yet the
shear and acceleration of the fluid flow are not dominated by its
expansion as the singularity is approached, as one would expect
with a Friedmann-like singularity. The fluid congruence is not
regular at the IS in the Coley-Tupper models, however, which
motivated Goode and \mbox{Wainwright \cite{GW85}} to include the
following additional definition relating to the fluid flow.

\begin{defn}[Fluid congruence]
\label{defn:fluid_congruence} With any unit timelike congruence
$\vec{u}$ in $\M$ we can associate a unit timelike congruence $\s
\vec{u}$ in $\s \M$ such that \be \s \vec{u} = \Omega \vec{u}
\quad \text{in }\:\M \, .
\end{equation}
\begin{itemize}
\item[(a)] If we can choose $\s \vec{u}$ to be regular
           (at least $C^3$) on
           an open neighbourhood of
           $T=0$ in $\s\M$, we say that $\vec{u}$ is
           \emph{regular at the isotropic singularity}. \\
\item[(b)] If, in addition, $\s \vec{u}$ is orthogonal to $T=0$, we say
that $\vec{u}$ is \emph{orthogonal to the isotropic singularity}.
\end{itemize}
\end{defn}

It is the requirement that the fluid flow be regular at an IS
(condition (a) of Definition \ref{defn:fluid_congruence}), which
ensures that the appropriate kinematic quantities behave as one
would expect as an ``isotropic'' singularity is approached
\cite{GW85}.  In the remainder of this \thisdocthesis
\hspace{-1.5mm}, unless stated otherwise, it will be assumed that
any space-time which admits an IS has a fluid flow which is
regular at the IS.

It should be noted that, in relation to Proposition
\ref{prop:conformal_factor_freedom}, if the fluid congruence
$\vec{u}$ is regular at an IS, then $\s\vec{u}=\Omega\vec{u}$ (on
$\M$) can be chosen to be at least $C^3$ on an open neighbourhood
of $T=0$ in $\s\M$.  Since $k(T)$ is a $C^3$ function of $T$ on $
\R$, then $\s\tilde{\vec{u}} = e^{k(T)}\s\vec{u}$ is also at least
$C^3$ on an open neighbourhood of $T=0$ in $\s\M$.  It follows
that $\vec{u}$ is also regular at the IS associated with the
second unphysical space-time $(\s\M,\s\tilde{\tns{g}})$.

In the following proposition we demonstrate how the freedom in the
choice of conformal factor associated with an IS (see Proposition
\ref{prop:conformal_factor_freedom}) can be used to ensure that
the unphysical expansion scalar $\s\theta$ is zero when evaluated
at a given point $p$ in $\s\M$ with $T\geq 0$.  Goode and
Wainwright \cite{GW85} have previously examined the case of an
irrotational barotropic perfect fluid model which admits an IS at
which the fluid flow is regular, and the fluid flow is orthogonal
to the family of spacelike hypersurfaces defined by $T=$ constant,
where $T$ is the cosmic time function, and shown that the
unphysical expansion scalar $\s\theta$ is constant (with zero as
one possible choice) on the $T=0$ hypersurface.  Proposition
\ref{prop:stheta_approaches_zero} shows that a conformal factor
can be chosen which sets $\s\theta=0$ everywhere on this
hypersurface.

\begin{prop}
\label{prop:stheta_approaches_zero} Suppose that the space-time
$(\M,\tns{g})$ admits an IS at which the fluid flow $\vec{u}$ is
regular.  Let $p$ be a point (with $T\geq 0$) in the conformally
related unphysical space-time $(\s\M,\s\tns{g})$.  Then by the
suitable choice of a new conformal factor $\tilde{\Omega}$ (see
Proposition \ref{prop:conformal_factor_freedom}) we can ensure
that \bea {\s \tilde{\theta}|}_p=0. \eea
\end{prop}

\begin{proof}
For this proof we will use the conformal structure detailed in the
proof of Proposition \ref{prop:conformal_factor_freedom}.

The expansion scalars $\theta$ and $\s\theta$ for the two
conformally related space-times $(\M,\tns{g})$ and
$(\s\M,\s\tns{g})$ are related by \be
\s\theta=\Omega\theta-3\frac{\Omega'}{\Omega}T_{,a}\s u^a \ .
\end{equation}
The expansion scalar $\s\tilde{\theta}$ for the second unphysical
space-time $(\s\M,\s\tilde{\tns{g}})$ is therefore related to the
expansion scalar $\s\theta$ for the first unphysical space-time
$(\s\M,\s\tns{g})$ by
\bea \s\tilde{\theta}&=&e^{k(T)}\s\theta-3k' T_{,a}\s \tilde{u}^a \\
&=& e^{k(T)}[\s\theta-3k'T_{,a}\s u^a],
\end{eqnarray}
where this relationship is valid on $\M$ (i.e., on $\s\M$ for
$T>0$), as well as on an open neighbourhood of $T=0$ in $\s\M$.
Thus, for $p\in\s\M$ (with $T\geq0$), we can arrange to have
$\s\tilde{\theta}|_p=0$ by choosing the function $k(T)$ such that
\be k'|_{T(p)}=\frac{1}{3}\frac{\s\theta|_p}{(T_{,a}\s{u}^a)|_p}\
,
\end{equation}
noting that $T_{,a}\s{u}^a\neq 0$.
\end{proof}


\section{Previous Results}
\label{sec:review} It is interesting to review the main physical
results which have been achieved using the amended version of the
Goode and Wainwright definition of an IS given by Scott. It is
fair to say that, until recently, the only major issue tackled
with great success in this field was the vorticity problem. Goode
\cite{Goode87} showed that if a perfect fluid solution of the EFE
with an exact gamma-law equation of state $(1 < \gamma < 2)$
admits an IS at which the fluid flow is regular, then the fluid
flow must be irrotational.  Scott \cite{SSGVR,Scott88,Scott89} has
extended this result to include all barotropic perfect fluids with
$-1 < \lambda < 1$. This result is known as the General Vorticity
Result (GVR).

As well as establishing the GVR, Scott \cite{SSGVR} examines in
detail, and amends, the definition of an IS, and also determines
the form of the conformal factor in terms of the cosmic time
function. Amongst other results, Scott provides a rigorous proof
that if a perfect fluid (not necessarily barotropic) solution of
the EFE admits an IS ($\lambda \not = -\infty$ or
$\smallfrac{1}{2}$) at which the fluid flow is regular, then there
exists a limiting gamma-law equation of state as the singularity
is approached. The limiting behaviour, as the singularity is
approached, of an acceleration potential for the fluid flow is
also established for barotropic perfect fluids in this category.

Various examples of cosmological models which admit an isotropic
singularity at which the fluid flow is regular are presented in
the literature. A discussion of these models forms a review paper
\cite{potsdam}, which also includes a classification of the models
according to the limiting behaviour, as the IS is approached, of
their various kinematical properties and tensorial quantities.

Recently the authors have established the Zero Acceleration Result
\cite{ZAR} (ZAR) for solutions of the EFE with barotropic perfect
fluid source. This result says that if the fluid congruence of
such a solution is shear-free and regular at an IS (with $-1 <
\lambda < 1$), then the fluid flow is necessarily geodesic.


\section{Vacuum cosmologies}
\label{sec:vacuum_cosmologies}

If a solution of the Einstein field equations admits an IS,
then it cannot be a vacuum solution\footnote{This
result agrees with the analogous result, by Lifshitz and
Khalatnikov \cite[p203]{LK63}, for the quasi-isotropic
singularity.}. This is easily seen by contrasting the fact that,
globally, a vacuum solution satisfies \be {R_a}^b{R^a}_b=0,
\end{equation}
with the result, by Goode and Wainwright \cite{GW85}, that as an
IS at T = 0 is approached, \be
\lim_{T\rightarrow0^+}{R_a}^b{R^a}_b=\infty.
\end{equation}
In light of this result we henceforth confine our attention to
fluid filled space-times.


\section{Singularity type}
\label{sec:singularity_type} In order to gain some insight into
the behaviour of the fluid around the singularity in a
cosmological model we take a fluid element, which at some finite
time is assumed to be spherical, and examine the asymptotic
behaviour of this fluid element as we approach the initial
singularity back along the fluid flow lines\footnote{This approach
was pioneered by Thorne \cite{Thorne67}, but this section is based
on MacCallum \cite[p131]{MacCallum73}.}.  To do this we look at
length scales $l_\alpha$ in the eigendirections\footnote{Note that
the eigenvectors of the expansion tensor are in the same
directions as the eigenvectors of the shear tensor.} of the
expansion tensor.  The length scales $l_\alpha$ $(\alpha = 1,2,3)$
are defined, up to a multiplicative constant, by \be
\frac{\dot{l}_\alpha}{l_\alpha}\equiv {\theta^\alpha}_{\alpha},
         \quad\text{$\alpha$ not summed},
\end{equation}
where ${\theta^\alpha}_{\alpha}$ are the components of the
expansion tensor in its eigenframe.  An overall length scale (also
called the scale factor) is defined by \be
l=(l_1l_2l_3)^{\frac{1}{3}}\quad\text{or equivalently by}
      \quad \frac{\dot{l}}{l} = \frac{\theta}{3}.
\end{equation}
A singularity is defined to be
\begin{itemize}
\item a \emph{point-like} singularity if all three $l_\alpha$
      approach zero,
\item a \emph{barrel} singularity if two of the $l_\alpha$ approach zero
      and the other approaches some finite number,
\item a \emph{cigar} singularity if two of the $l_\alpha$ approach zero
      and the other approaches infinity,
\item a \emph{pancake} singularity if one of the $l_\alpha$ approaches
      zero and the other two approach some finite number.
\end{itemize}
It should be noted that it is possible for a singularity to be
none of the above singularity types.  For example, the mixmaster
cosmological models have the $l_\alpha$ oscillating as the
singularity is approached.

From the above choices of singularity type, it would be natural to
expect that the IS is a point-like singularity.  Indeed, this is
alluded to by Tod in a review paper \cite{Tod92} and proven for
the case of a geodesic flow by Nolan \cite{Nolan01}.

Before we proceed to prove, in general, that the IS actually is a
point-like singularity, we must first establish the following
result.

\begin{prop}
\label{prop:eigenvectors_of_physical_expansion_equal_unphysical_ones}
Suppose that the space-time $(\M,\tns{g})$ with a $C^1$ unit
timelike congruence $\tns{u}$ admits an IS. We denote the
corresponding $C^1$ unit timelike congruence in the unphysical
space-time $(\s\M,\s\tns{g})$ by $\s\vec{u} = \Omega \vec{u}$. For
$p\in \M$, a vector, $\vec{v} \in T_p(\M)$, where $\vec{v}$ is
orthogonal to $\vec{u}|_p$, is an eigenvector (with eigenvalue
$\kappa$) of the expansion tensor, ${\theta^a}_b$, of $\vec{u}$ if
and only if it is an eigenvector (with eigenvalue $\s\kappa$) of
the expansion tensor, $\s{\theta^a}_b$, of $\s\vec{u}$.
\end{prop}
\begin{proof}
We note firstly that since $\vec{u}$ is orthogonal to ${\theta^a}_b$
and $\s{\theta^a}_b$, then $\vec{u}$ is an eigenvector of both
${\theta^a}_b$ and $\s{\theta^a}_b$ with zero eigenvalues.  That
is, the eigenvalues $\kappa$ and $\s\kappa$ are both zero for
the eigenvector $\vec{u}$.

Now for $p\in\M$, coordinates $(x^a)$ about $p$ can be
chosen so that $\frac{\partial\ \ }{\partial x^0}|_p = \vec{u}|_p$
and $\tns{g}(\frac{\partial\ \ }{\partial x^0},\frac{\partial\ \
}{\partial x^\alpha})|_p = 0 \ (\alpha = 1,2,3)$.  It follows from
this that $\s\tns{g}(\frac{\partial\ \ }{\partial
x^0},\frac{\partial\ \ }{\partial x^\alpha})|_p =\s\tns{g}(\s
\vec{u},\frac{\partial\ \ }{\partial x^\alpha})|_p = 0$.

It is readily seen from the form of ${\theta^a}_b$ and
$\s{\theta^a}_b$ in these coordinates, that any eigenvectors of
${\theta^a}_b$ at $p$, with non-zero eigenvalues, are orthogonal to
$\vec{u}$, and any eigenvectors of $\s{\theta^a}_b$ at $p$, with
non-zero eigenvalues, are orthogonal to $\s\vec{u}$.

Now consider the vector $\beta\vec{u} +\vec{w} \in T_p\M$, where
$\beta\neq 0$ and $\vec{w} \in T_p\M$ is orthogonal to $\vec{u}$.
If $\beta\vec{u} +\vec{w}$ is an eigenvector of ${\theta^a}_b$ at
$p$ with zero eigenvalue, then \bea
{\theta^a}_b (\beta u^b +w^b) &=& 0 \\
\iff\qquad\qquad {\theta^a}_b w^b &=& 0 \eea That is, $\vec{w}$ is
also an eigenvector of ${\theta^a}_b$ at $p$ with zero eigenvalue.

We recall that the conformal relationship between the expansion
tensor, ${\theta^a}_b$, of $\vec{u}$ and the expansion tensor,
$\s{\theta^a}_b$, of $\s\vec{u}$ is given by \be
\Omega{\theta^a}_b = \s{\theta^a}_b +\s{h^a}_b \Omega^{-1}
\Omega_{,d}\s u^d \ .
\end{equation}

Now for $p\in\M$, suppose that $\vec{v}$ is an eigenvector of
${\theta^a}_b$ at $p$ which is orthogonal to $\vec{u}$ : \bea
&\text{i.e.} &{\theta^a}_b v^b = \kappa v^a  \\
&\iff& \Omega{\theta^a}_b v^b = \Omega\kappa v^a \qquad(\Omega>0)\\
&\iff& \s{\theta^a}_b v^b +\s{h^a}_b v^b \Omega^{-1}\Omega_{,d}\s
u^d
   = \Omega\kappa v^a \\
&\iff& \s{\theta^a}_b v^b = (\Omega \kappa -
\Omega^{-1}\Omega_{,d}\s u^d)v^a
\quad (\text{since } \s{h^a}_b v^b = v^a ) \\
&\iff& \s{\theta^a}_b v^b = \s\kappa v^a \qquad \text{where }
\s\kappa =\Omega \kappa - \Omega^{-1}\Omega_{,d}\s u^d
\end{eqnarray}
i.e., $\vec{v}$ is an eigenvector of $\s{\theta^a}_b$ at $p$ which
is orthogonal to $\s \vec{u}$.
\end{proof}

\begin{thm}
\label{thm:IS_has_point-like_singularity}  If the space-time
$(\M,\tns{g})$ admits an IS at which the fluid flow is regular,
then $(\M,\tns{g})$ has a point-like singularity.
\end{thm}

\begin{proof}
We recall that the expansion tensor, ${\theta^a}_b$, of $\vec{u}$
for the physical space-time $(\M,\tns{g})$ is related to its
unphysical equivalent, $\s{\theta^a}_b$, of $\s\vec{u}$ for the
unphysical space-time $(\s\M,\s\tns{g})$ by the equation \bea
{\theta^a}_b &=&
     \Omega^{-1} {\s \theta^a}_b
      +\Omega^{-1}{\s h^a}_b{(\ln \Omega)}_{,d}{\s u}^{d} .
\label{eqn:relate_physical_unphysical_expansion} \eea

Choose an arbitrary flow line in $\s\M$ and label it $\gamma$. The
point $p\in\gamma$ lies on the IS hypersurface at $T=0$. We now
proceed to construct a local comoving coordinate system
$(s,x^\alpha)$ in $(\s\M,\s\tns{g})$.  Set $s=zT$ along $\gamma$,
where $T$ is the cosmic time function defined on $\s\M$, and
$z\in\R^+$ is a constant. The coordinate $s$ will be strictly
monotonically increasing up $\gamma$.  Now consider a small open
tube $\Nbd$ of flow lines about $\gamma$ in $\s\M$ and the
hypersurfaces, within $\Nbd$, which are orthogonal to the flow
lines. The coordinate $s$ will be chosen to be constant across
these orthogonal hypersurfaces. Note that the $s=$ constant
hypersurfaces are not necessarily the same as the $T=$ constant
hypersurfaces.  The constant $z$ is chosen so that
$\frac{\partial}{\partial s}|_p=\s\vec{u}|_p$. On the hypersurface
$s=0$ we choose spatial coordinates $x^\alpha$ which are comoving,
i.e., the spatial coordinates along any flow-line remain constant.

One further restriction is placed on the spatial coordinates. They
are chosen so that, at the point $p$ on the hypersurface $T=0$,
the coordinate tangent vectors $\frac{\partial}{\partial
x^\alpha}$ coincide with the three linearly independent
eigenvectors, ${}_\alpha\!\vec{v}$, at $p$, of the expansion
tensor $\s{\theta^a}_b$ of the unphysical fluid flow, $\s\vec{u}$.
The eigenvectors, ${}_\alpha\!\vec{v}$, are orthogonal to
$\s\vec{u}|_p$. The eigenvectors $\s\vec{u}, {}_\alpha\!\vec{v}$
of $\s{\theta^a}_b$ at $p$ can be extended to form a $C^2$
eigenvector frame $(\s\vec{u}, {}_\alpha\!\vec{v})$ on $\Nbd$,
where the eigenvectors ${}_\alpha\!\vec{v}$ are always orthogonal
to $\s\vec{u}$ at any point of $\Nbd$.   For points in $\Nbd$,
other than $p$, one would expect that ${}_\alpha\!\vec{v}$ will
not, in general, coincide with $\frac{\partial}{\partial
x^\alpha}$, although both ${}_\alpha\!\vec{v}$ and
$\frac{\partial}{\partial x^\alpha}$ are orthogonal to $\s
\vec{u}$. It is certainly true, however, that, along $\gamma$ \be
\s u^0 = 1+o(1), \quad \s u^\beta = 0,\quad {}_\alpha\! v^0 =
0,\quad {}_\alpha\! v^\beta = {\delta^\beta}_\alpha +
o(1)\quad\text{as }s\rightarrow 0^+.
\end{equation}

The coordinate tangent vector frame is thus arbitrarily close to
the expansion tensor $\s{\theta^a}_b$ eigenvector frame as we
approach $p$ down the flow line $\gamma$.  Since, by Proposition
\ref{prop:eigenvectors_of_physical_expansion_equal_unphysical_ones},
the expansion tensor $\s{\theta^a}_b$ eigenvector frame is the
same as the expansion tensor ${\theta^a}_b$ eigenvector frame, it
follows that the coordinate tangent vector frame is also
arbitrarily close to the expansion tensor ${\theta^a}_b$
eigenvector frame as we approach $p$ down the flow line $\gamma$.
A leading subscript ``e'' will be placed on any tensor evaluated
in the expansion tensor ${\theta^a}_b$ eigenvector frame.

 We will now consider the relationship between the
components of the expansion tensor ${\theta^a}_b$ in its
eigenvector frame and its components in the coordinate tangent
vector frame.  For $\alpha = 1,2,3$, Equation
(\ref{eqn:relate_physical_unphysical_expansion}) can be written
\be \Omega{\theta^\alpha}_\alpha = \s{\theta^\alpha}_\alpha +
\s{h^\alpha}_\alpha\frac{\Omega'}{\Omega}(\s\nabla T)_d\s u^d
,\quad\alpha\text{ not summed.}
\label{eqn:physical_and_unphysical_theta_diagonal_terms}
\end{equation}
The corresponding equation in the eigenvector frame is \be
\Omega{}_{e}{\theta^\alpha}_\alpha =
{}_{e}\!\!\s{\theta^\alpha}_\alpha
+{}_{e}\!\!\s{h^\alpha}_\alpha\frac{\Omega'}{\Omega}{}_{e}\!(\s\nabla
T)_d\,{}_e\!\!\s u^d \ .
\end{equation}
As we approach $p$ down the flow line $\gamma$ in
$\s\M$, \bea {}_{e}\!\!{\s\theta^\alpha}_\alpha &=&
{\s\theta^\alpha}_\alpha [1+o(1)],
\\{}_{e}\!\!\s {h^\alpha}_\alpha &=& \s{h^\alpha}_\alpha[1+o(1)], \\
{}_{e}\!(\s\nabla T)_d &=&(\s\nabla T)_d[1+o(1)], \\
{}_e\!\!\s u^d &=& \s u^d [1+o(1)]. \eea It follows that \be
\Omega {}_{e}{\theta^\alpha}_\alpha = \Omega
{\theta^\alpha}_\alpha[1+o(1)] \quad\text{as }s\rightarrow 0^+,
\end{equation}
and thus \be {}_{e}{\theta^\alpha}_\alpha =
{\theta^\alpha}_\alpha[1+o(1)] \quad\text{as }s\rightarrow 0^+.
\end{equation}

In the physical expansion tensor, ${\theta^a}_b$, eigenvector
frame the length scales, $l_\alpha\ (\alpha = 1,2,3)$, are defined
by $\frac{\dot{l}_\alpha}{l_\alpha} =
{}_{e}{\theta^\alpha}_\alpha$, where $\alpha$ is not summed. We
therefore have the relationship \be
\frac{\dot{l}_\alpha}{l_\alpha} = {\theta^\alpha}_\alpha
[1+o(1)]\quad\text{as }s\rightarrow 0^+,
\end{equation}
which in conjunction with Equation
(\ref{eqn:physical_and_unphysical_theta_diagonal_terms}) yields
\bea \Omega \frac{\dot{l}_\alpha}{l_\alpha} &=& \left(
\s{\theta^\alpha}_\alpha
 + \s{h^\alpha}_\alpha\frac{\Omega'}{\Omega}(\s\nabla T)_d\s u^d
   \right)[1+o(1)] \ .
\eea Now as we approach $p$ down the flow line $\gamma$ in $\s\M$,
$\s{\theta^\alpha}_\alpha|_p \in\R$, $\lim_{s\rightarrow
0^+}\frac{\Omega'}{\Omega}=+\infty$ (see Scott \cite{SSGVR}), and
\be\lim_{s\rightarrow 0^+}\s{h^\alpha}_\alpha(\s\nabla T)_d\s u^d
= \lim_{s\rightarrow 0^+}(\s\nabla T)_d\s u^d \in
\R\backslash\{0\}.\end{equation} It follows that \bea
 \Omega \frac{\dot{l}_\alpha}{l_\alpha} &=&
 \frac{\Omega'}{\Omega}(\s\nabla T)_d\s u^d
  \ [1+o(1)]
  \quad\text{as }s\rightarrow 0^+\label{eqn:aymptotic_ln_l_dot}\\
\Leftrightarrow\qquad\quad \Omega \frac{\dot{l}_\alpha}{l_\alpha}
&=& \Omega\frac{\Omega'}{\Omega}(\nabla T)_d u^d
  \ [1+o(1)]   \\
 \Leftrightarrow\quad\,\, \Omega(\ln
l_\alpha)\dot{} &=& \Omega(\ln
\Omega)\dot{}\ [1+o(1)]\\
\Leftrightarrow\qquad (\ln l_\alpha)\dot{} &=& (\ln \Omega)\dot{}\
[1+o(1)]\ . \label{eqn:dot_ln_l_approx_dot_ln_omega}\eea Thus
$\lim_{s\rightarrow 0^+}\frac{(\ln l_\alpha)\dot{}}{(\ln
\Omega)\dot{}}=1$.

Since $\ln \Omega\rightarrow -\infty$ as $s\rightarrow 0^+$,
L'H\^{o}pital's rule can be invoked to obtain \bea
\lim_{s\rightarrow 0^+}\frac{\ln l_\alpha}{\ln
\Omega}&=&1 \\
\text{i.e.}\qquad\qquad\quad \ln l_\alpha &=& \ln\Omega\ [1+o(1)]
\quad\text{as
}s\rightarrow 0^+\\
\implies\qquad\qquad\qquad  l_\alpha &\rightarrow& 0\quad\text{as
}s\rightarrow 0^+.\eea

Note that the length scales, $l_\alpha \ (\alpha=1,2,3)$, have the
same asymptotic form, as one would expect with an IS. Since
$\gamma$ was an arbitrary flow line, we conclude that all three
length scales $l_\alpha$ approach zero as the initial singularity
is approached back along any fluid flow line.  Thus the space-time
$(\M,\tns{g})$ has a point-like singularity.

\end{proof}

This theorem shows that if a cosmological model admits an IS at
which the fluid flow is regular, then a spherical fluid element
which travels backwards in time to the initial singularity, will
collapse, approximately isotropically, into a point.


\section{Hubble parameter}
\label{sec:hubble_parameter}

The Hubble parameter is an important parameter in comparing
observational data with cosmological theory.  It is defined by
MacCallum \cite[p97]{MacCallum73} as \be H \equiv
\frac{\dot{l}}{l} = \frac{\theta}{3},
\end{equation}
where $l$ is the scale factor.

The behaviour of the expansion scalar, $\theta$, can be determined
as follows. Equation (\ref{eqn:aymptotic_ln_l_dot}), from the
proof of Theorem \ref{thm:IS_has_point-like_singularity}, shows
that for a cosmological model which admits an IS, at which the
fluid flow is regular, the following relationship holds down a
fluid flow line: \be
 \frac{\dot{l}_\alpha}{l_\alpha} =
 \frac{1}{\Omega}\frac{\Omega'}{\Omega}(\s\nabla T)_d\s u^d
  \ [1+o(1)]\quad\text{as }T\rightarrow 0^+.
\end{equation}
Since $\lim_{T\rightarrow 0^+}\frac{\Omega'}{\Omega^2}=+\infty$
and $\lim_{T\rightarrow 0^+}(\s\nabla T)_d\s u^d \in \R^+$, it
follows that \be \frac{\dot{l}_\alpha}{l_\alpha} \rightarrow
+\infty\quad\text{as } T\rightarrow 0^+.
\end{equation}
Now \bea
\theta &=& \frac{\dot{l}_1}{l_1} + \frac{\dot{l}_2}{l_2}
+ \frac{\dot{l}_3}{l_3} \\
\implies\quad \theta &\rightarrow& +\infty \quad\text{as }
T\rightarrow 0^+.
\end{eqnarray}
This result was originally obtained by Goode and Wainwright
\cite{GW85}.

Since the expansion scalar becomes infinite as the singularity is
approached along any fluid flow line, it follows that \be H
\rightarrow +\infty \quad\text{as } T \rightarrow 0^+.
\end{equation}


\section{Energy conditions}
\label{sec:energy_conditions}

A solution of the EFE should be ``physically reasonable'' in order
to be a candidate as a model for a real physical system.  Although
the terminology ``physically reasonable'' is vague, the standard
method for determining whether or not a model is physically
reasonable is to test if the energy conditions are satisfied. For
a perfect fluid the usual energy conditions are
\cite[pp88-96]{HE73}:
\begin{description}
\item[Weak Energy Condition (WEC):] $\quad\mu\geq 0,\quad\mu+p\geq 0$.
\item[Dominant Energy Condition (DEC):]$\quad\mu\geq 0,
                                          \quad -\mu\leq p\leq \mu$.
\item[Strong Energy Condition (SEC):]$\quad\mu+p\geq 0,\quad\mu+3p\geq 0$.
\end{description}

Scott \cite{SSGVR} has shown that if a
space-time satisfies the EFE with perfect fluid source, and the
unit timelike fluid congruence is regular at an IS, then there is
a relationship between the energy density, $\mu$, and the
pressure, $p$, of the fluid near the singularity. The following
two propositions give this relationship precisely.

\begin{prop}[Limiting $\gamma$-law result]
If the space-time $(\M,\tns{g})$ is a $C^3$ solution of the EFE
with perfect fluid source, and the unit timelike fluid congruence,
$\tns{u}$, is regular at an IS ($\lambda\neq -\infty $ or
$\smallfrac{1}{2}$), then there exists a limiting $\gamma$-law
equation of state $p=(\gamma-1)\mu$ as the singularity is
approached, where $\gamma=\smallfrac{2}{3}(2-\lambda)$.
\end{prop}
\begin{prop}[$\lambda = -\infty$]
If the space-time $(\M,\tns{g})$ is a $C^3$ solution of the EFE
with perfect fluid source, and the unit timelike fluid congruence,
$\tns{u}$, is regular at an IS with $\lambda=-\infty$, then there
exists a limiting equation of state $p= -\frac{2}{3}L\mu$ as the
singularity is approached, where $L(T)=
\frac{\Omega''\Omega}{(\Omega')^2}$, and $\lim_{T\rightarrow
0^+}L(T) = \lambda.$
\end{prop}

Using these results, we have determined, in the following
corollary, whether, for perfect fluid cosmologies which admit an
IS at which the fluid flow is regular, the above energy conditions
are satisfied near the singularity.

\begin{cor}
If the space-time $(\M,\tns{g})$ is a $C^3$ solution of the EFE
with perfect fluid source, and the unit timelike fluid congruence,
\vec{u}, is regular at an Isotropic Singularity, then there exists
an open neighbourhood $\U$ of the hypersurface $T=0$ in $\s\M$
such that the weak and strong energy conditions are satisfied
everywhere on $\U\cap\M$. Furthermore, if $-1<\lambda<1$, then the
dominant energy condition also holds on $\U\cap\M$.
\end{cor}

\begin{proof}
Scott \cite{SSGVR} has
shown that, for space-times
which satisfy the conditions of this corollary, $\mu\rightarrow
+\infty$ as $T\rightarrow 0^+$, and hence $\mu\geq 0$ near the
singularity.

From the limiting $\gamma$-law result of Scott
we know that when $\lambda\neq -\infty$
or $\smallfrac{1}{2}$, $p\approx (\gamma -1)\mu$ as $T\rightarrow
0^+$, with $\smallfrac{2}{3}<\gamma<\infty \ (\gamma\neq 1)$ .  We
will prove this case first, then look at the remaining
$\lambda=\smallfrac{1}{2}$ and $\lambda=-\infty$ cases.
\begin{enumerate}   
    \item   \label{item:energy_conds_lambda_neq_infty}
    Assume $\lambda\neq -\infty$ or $\smallfrac{1}{2}$.

    An
    open neighbourhood $\U$ of the hypersurface $T=0$ in $\s\M$
    can be chosen such that $\mu>0$ and $p\approx(\gamma -1)\mu$
    as $T\rightarrow 0^+$ on $\U\cap\M$.

    There are three subcases to examine.
    \begin{enumerate}   
        \item
        \label{item:mu_plus_p_greater_than_zero_normal}
        Show $\mu+p\geq0$  on $\U\cap\M$:
            \bea
            \mu+p   &=& \mu+(\gamma -1)\mu[1+o(1)] \\
                &=& \gamma\mu[1+o(1)]\  .
            \eea
            Now $\gamma>\smallfrac{2}{3}$ and
            $\mu>0$ on $\U\cap\M$, with $\mu\rightarrow +\infty$
            as $T\rightarrow 0^+$. Hence $\mu+p \geq 0$ on
            $\U\cap\M$ and the WEC is satisfied.
        \item   Show $\mu+3p\geq 0$ on $\U\cap\M$:

            Proceeding as in
            case (\ref{item:mu_plus_p_greater_than_zero_normal}),
            we can show that
            \be
            \mu+3p=(3\gamma-2)\mu[1+o(1)]\ ,
            \end{equation}
            and since $\gamma>\smallfrac{2}{3}$, $3\gamma-2>0$.
            Hence $\mu+3p \geq 0$ on $\U\cap\M$ and the SEC is
            satisfied.
        \item Establish when $-\mu\leq p\leq\mu$ holds on $\U\cap\M$:

            We have already shown that $-\mu\leq p$ on $\U\cap\M$,
            and so we need to
            determine when it is also true that $p\leq\mu$.

       Since $p=(\gamma-1)\mu[1+o(1)]$ as $T\rightarrow0^+$ on $\U\cap\M$,
            \bea
            & &p\leq\mu \\
            &\lra& (\gamma-1)\mu[1+o(1)] \leq \mu \\
            &\lra& (\gamma-1)[1+o(1)] \leq 1 \\
       & &\quad(\text{since } \mu > 0 \text{ on } \U\cap\M).\nonumber
            \eea
            When $\gamma-1<1$, an open neighbourhood $\U$ of $T=0$
            in $\s\M$ always exists such that this inequality
            holds on $\U\cap\M$.  On the other hand, when
            $\gamma-1>1$, such a neighbourhood never exists, since
            the inequality is not true in the limit as
            $T\rightarrow0^+$.
              For the
       special value $\gamma=2$, the inequality may or may not hold
       in the limit as $T\rightarrow0^+$.  We
       conclude that, for $\lambda\neq -\infty$ or $\smallfrac{1}{2}$,
       the DEC holds on $\U\cap\M$ when $\smallfrac{2}{3}< \gamma <2
       \ (\gamma\neq1)$,
       equivalently, $-1<\lambda<1\ (\lambda\neq\smallfrac{1}{2})$.
        \end{enumerate}     

    \item   Assume $\lambda = \smallfrac{1}{2}$.

     This case includes the dust $(p=0)$ and asymptotic dust ($p=o(1)$
     as $T\rightarrow 0^+$) models.
        It is readily seen from \cite{SSGVR} that $p=o(\mu)$
        as $T\rightarrow0^+$.  Let $\U$ denote an open neighbourhood
        of the hypersurface $T=0$ in $\s\M$ such that $\mu>0$
         and $p=o(\mu)$ as $T\rightarrow 0^+$ on  $\U\cap\M$.
         Since
     $\mu\rightarrow +\infty$ as $T\rightarrow 0^+$, it is clear
     that $\mu+p \geq 0$, $\mu+3p \geq 0$, and $p\leq\mu$ on $\U\cap\M$.
     Thus, for the case $\lambda = \smallfrac{1}{2}$, the WEC, SEC
     and DEC are satisfied on $\U\cap\M$.

    \item   Assume $\lambda=-\infty$.

        In this case the limiting equation of state is given by
     \be
     p\approx -\frac{2}{3}L\mu \quad\text{as }T\rightarrow 0^+,
     \end{equation}
     where $L(T)= \frac{\Omega''\Omega}{(\Omega')^2}$,
     and $\lim_{T\rightarrow 0^+}L(T) = \lambda.$  An open
     neighbourhood $\U$ of the hypersurface $T=0$ in $\s\M$ can be
     chosen such that $\mu>0$, $L<0$, and
     $p\approx-\frac{2}{3}L\mu$ as $T\rightarrow 0^+$  on
     $\U\cap\M$.

        Again there are three subcases to examine.
        \begin{enumerate} 
        \item   \label{item:mu_plus_p_greater_than_zero_infinity}
        Show $\mu+p\geq0$ on $\U\cap\M$:
            \bea
            \mu+p &=& \mu-\frac{2}{3}L\mu[1 + o(1)] \\
                  &=& -\smallfrac{2}{3}L\mu[1 + o(1)]\ .
            \eea
            Now $\mu>0$ and $L<0$ on $\U\cap\M$, with
            $-\frac{2}{3}L\mu\rightarrow +\infty$
            as $T\rightarrow 0^+$. Hence $\mu+p \geq 0$
            on $\U\cap\M$ and the WEC is satisfied.
        \item   Show $\mu+3p\geq 0$ on $\U\cap\M$:

       Proceeding as in
       case (\ref{item:mu_plus_p_greater_than_zero_infinity}),
       we can show that
            \be
            \mu+3p=-2L\mu[1+o(1)].
            \end{equation}
            Now $\mu>0$ and $L<0$ on $\U\cap\M$, with
            $-2L\mu\rightarrow +\infty$
            as $T\rightarrow 0^+$.
            Hence $\mu+3p \geq 0$ on $\U\cap\M$ and the SEC is satisfied.
        \item   Establish whether $-\mu\leq p\leq\mu$ holds on $\U\cap\M$:

            We have already shown that $-\mu\leq p$ on $\U\cap\M$,
            and so we need to
            determine if it is also true that $p\leq\mu$.

       Since $p=-\smallfrac{2}{3}L\mu[1+o(1)]$ as
       $T\rightarrow0^+$
       on $\U\cap\M$,
            \bea
            & &p\leq\mu \\
            &\lra& -\smallfrac{2}{3}L\mu[1+o(1)]\leq \mu \\
            &\lra& -\smallfrac{2}{3}L[1+o(1)]\leq 1\\
       & &\quad(\text{since }\mu>0\text{ on }
             \U\cap\M)\nonumber\\
            &\lra& 1+o(1)\leq -\frac{3}{2}\frac{1}{L}.
            \eea
            An open neighbourhood $\U$ of $T=0$ in $\s\M$ does not
            exist such that this inequality holds on $\U\cap\M$,
            since
       $-\smallfrac{3}{2}\frac{1}{L}\rightarrow 0^+$
       as $T\rightarrow0^+$.
       We conclude that the DEC does not hold when
       $\lambda = -\infty$, which is
       what one would expect given our results for the general
       case (\ref{item:energy_conds_lambda_neq_infty}) when
       $\lambda \neq -\infty$ or $\smallfrac{1}{2}$.
        \end{enumerate}
\end{enumerate} 
\end{proof}

The dominant energy condition may hold when $\gamma=2$, but this
must usually be examined on a case by case basis. We can, however,
state that any perfect fluid space-time with the exact
$\gamma$-law equation of state, $p=\mu$, which admits an IS at
which the fluid flow is regular, will satisfy the DEC.  For
example, the stiff fluid FRW models, and the stiff fluid Mars95
models \cite{Mars95}, admit an IS at which the fluid flow is
regular \cite[and references therein]{potsdam} and satisfy the
DEC. On the other hand, perfect fluid models with an equation of
state $p=\mu[1+N]$, where $N=o(1)$ as $T\rightarrow 0^+$, $N>0$,
which admit an IS at which the fluid flow is regular, do not
satisfy the DEC.


\section{Conclusion}
\label{sec:prop_conclusion}

As in any field, there is a certain amount of ``folk lore'' in
general relativity.  In particular, with regard to initial
singularities in cosmological models, it has always been assumed
that the IS is a point-like singularity and that vacuum
space-times were of no concern in matters regarding the IS, in
that they could not possess an IS. In this \thisdoc we have
explicitly proven these two fundamental results.  In addition to
this we have shown that, near the initial singularity, the weak
and strong energy conditions are automatically satisfied by
perfect fluid space-times which admit an IS, and that the dominant
energy condition is also satisfied given a reasonable restriction
on the relationship between the pressure and the energy density.
We have provided a detailed proof of the result,
formerly stated by Goode and Wainwright, that there is a freedom
in the choice of conformal factor associated with a space-time
which admits an IS. This freedom was exploited to prove that the
expansion scalar associated with an unphysical fluid flow can
always be set to zero at a given point in $\s\M$ by a suitable
choice of the conformal factor. Finally, it was shown that the
Hubble parameter becomes infinite at an IS.




\newcommand{\journal}[7]{#1  ``#2''  \textit{#3} \textbf{#4} (#5) pp#6-#7}

\newcommand{\proc}[8]{#1  ``#2''  \textit{#3}  ed. #4 (#5) (#6) pp#7-#8}

\newcommand{\book}[4]{#1, \textit{#2} (#3, #4).}

\newcommand{\phd}[3]{#1 \textit{Ph.D. thesis} #2 (#3)}

\newcommand{\preprint}[4]{#1, \textit{#3} (#4)}

\newcommand{\AOP}{A.P.\ }
\newcommand{\AJ}{Astrophys.\ J.\ }
\newcommand{\AP}{Adv.\ Phys.\ }
\newcommand{\CMP}{Commun.\ math.\ Phys.\ }
\newcommand{\CQG}{Class.\ Quantum Grav.\ }
\newcommand{\GRG}{Gen.\ Rel.\ Grav.\ }
\newcommand{\JMP}{J.\ Math.\ Phys.\ }
\newcommand{\MNRAS}{Mon. Not. R. astr. Soc.}
\newcommand{\PL}{Phys. Lett.}
\newcommand{\PR}{Phys. Rev.}
\newcommand{\PRD}{Phys. Rev. D}
\newcommand{\PRSLA}{Proc. R. Soc. Lond. A}



\end{document}